\documentclass[pdftex,cite,aps,prl,superscriptaddress,amsmath,amsfonts,amssymb,twocolumn]{revtex4}

\renewcommand{\vec}[1]{\mathbf{#1}}
\usepackage[pdftex]{graphicx}
\usepackage{braket}
\usepackage{wasysym}
\usepackage{setspace}
\usepackage{color}

\newcommand{\figref}[1]{Fig.~\ref{fig:#1}}

\renewcommand{\eqref}[1]{Eq.~(\ref{eq:#1})}

\newcommand{\citeasnoun}[1]{Ref.~\onlinecite{#1}}

\newcommand{\add}[1]{\if\a\b{{\color{black} #1}}\else{#1}\fi}
\newcommand{\comm}[1]{\if\a\b{{\color{blue}\{\small \sc #1\}}}\else{}\fi}
\newcommand{\del}[1]{{\if\a\b{{\color{DarkGreen}[[#1]]}}\else{}\fi}}

\begin{document}

\title{Repulsive, nonmonotonic Casimir forces in a glide-symmetric
geometry}

\author{Alejandro W. Rodriguez}
\affiliation{Department of Physics,
Massachusetts Institute of Technology, Cambridge, MA 02139}
\author{J. D. Joannopoulos}
\affiliation{Department of Physics,
Massachusetts Institute of Technology, Cambridge, MA 02139}
\author{Steven G. Johnson}
\affiliation{Department of Mathematics,
Massachusetts Institute of Technology, Cambridge, MA 02139}

\begin{abstract}
  We describe a three-dimensional geometry that exhibits a repulsive
  Casimir force using ordinary metallic materials, as computed via an
  exact numerical method (no uncontrolled approximations).  The
  geometry consists of a zipper-like, glide-symmetric structure formed
  of interleaved metal brackets attached to parallel plates.
  Depending on the separation, the perpendicular force between the
  plates/brackets varies from attractive (large separations) to
  repulsive (intermediate distances) and back to attractive (close
  separations), with one point of stable equilibrium in the
  perpendicular direction.  This geometry was motivated by a simple
  intuition of attractive interactions between surfaces, and so we
  also consider how a rough proximity-force approximation of pairwise
  attractions compares to the exact calculations.
\end{abstract}

\maketitle

\begin{figure}[t]
\includegraphics[width=1.0\columnwidth]{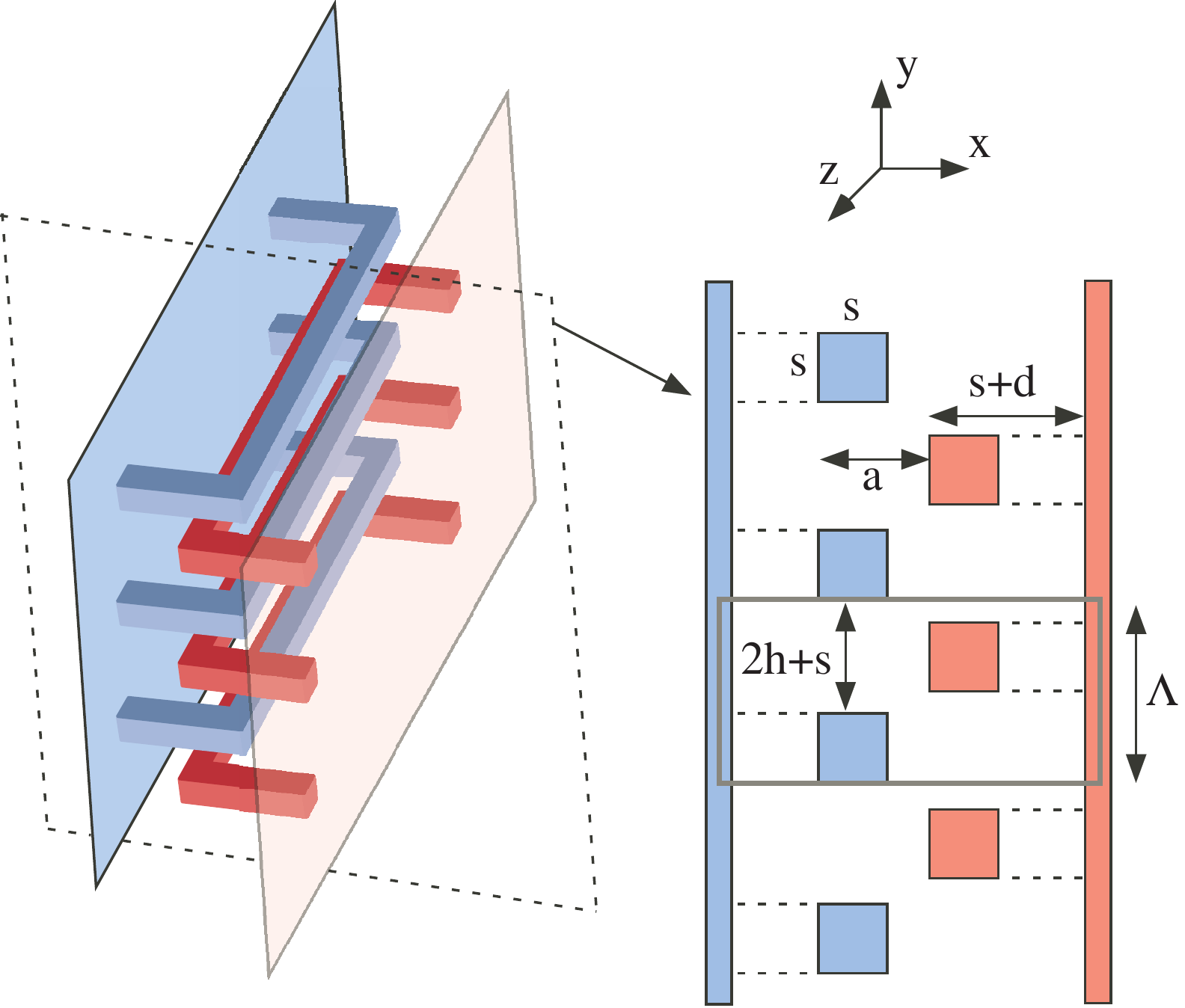}
\caption{(Color) Three-dimensional schematic of the Casimir ``zipper''
geometry of interlocking metal brackets (shown in different colors for
illustration only), along with a two-dimensional $xy$
cross-section. The dashed lines extruding from the plates to the
squares indicate their out-of-plane connectivity.}
\label{fig:geom}
\end{figure}

In this letter, we describe a metallic, glide-symmetric, ``Casimir
zipper'' structure (depicted in \figref{geom}) in which both repulsive
and attractive Casimir forces arise, including a point of stable
equilibrium with respect to perpendicular displacements. We compute
the force using an ``exact'' computational method (i.e. with no
uncontrolled approximations, so that it yields arbitrary accuracy
given sufficient computational resources), and compare these results
to the predictions of an {\it ad hoc} attractive interaction based on
the proximity-force approximation (PFA).  Casimir forces, a result of
quantum vacuum fluctuations, arise between uncharged objects, most
typically as an attractive force between parallel metal
plates~\cite{casimir} that has been confirmed
experimentally~\cite{moh3, decca2}.  One interesting question has been
whether the Casimir force can manifest itself in ways very different
from this monotonically decaying attractive force, and especially
under what circumstances the force can become repulsive.  It has been
proven that the Casimir force is always attractive in a
mirror-symmetric geometry (with $\varepsilon \geq 1$ on the
imaginary-frequency axis)~\cite{KennethKl06}, but there remains the
possibility of repulsive forces in asymmetric structures.  For
example, repulsive forces arise in exotic asymmetric material systems,
such as a combination of magnetic and electric
materials~\cite{Boyer74, ShaoZh06, Kenneth02}, fluid-separated
dielectric plates~\cite{Munday07}, metamaterials with
gain~\cite{Leonhardt07}, or excited atoms~\cite{Sherkunov05}.  Another
route to unusual Casimir phenomena is to use conventional materials in
complex geometries, which have been shown to enable asymmetrical
lateral ``ratchet'' effects~\cite{Emig07:ratchet} and nonmonotonic
dependencies on external parameters~\cite{Rodriguez07:PRL}.  Until
recently, however, predictions of Casimir forces in geometries very
different from parallel plates have been hampered by the lack of
theoretical tools capable of describing arbitrary geometries, but this
difficulty has been addressed by recent numerical
methods~\cite{Gies03, emig06, Rodriguez07:PRA, Emig07}.  In this
letter, we use a technique based on the mean Maxwell stress tensor
computed numerically via an imaginary-frequency Green's function,
which can handle arbitrary geometries and
materials~\cite{Rodriguez07:PRA}.


\begin{figure}[t]
\includegraphics[width=1.0\columnwidth]{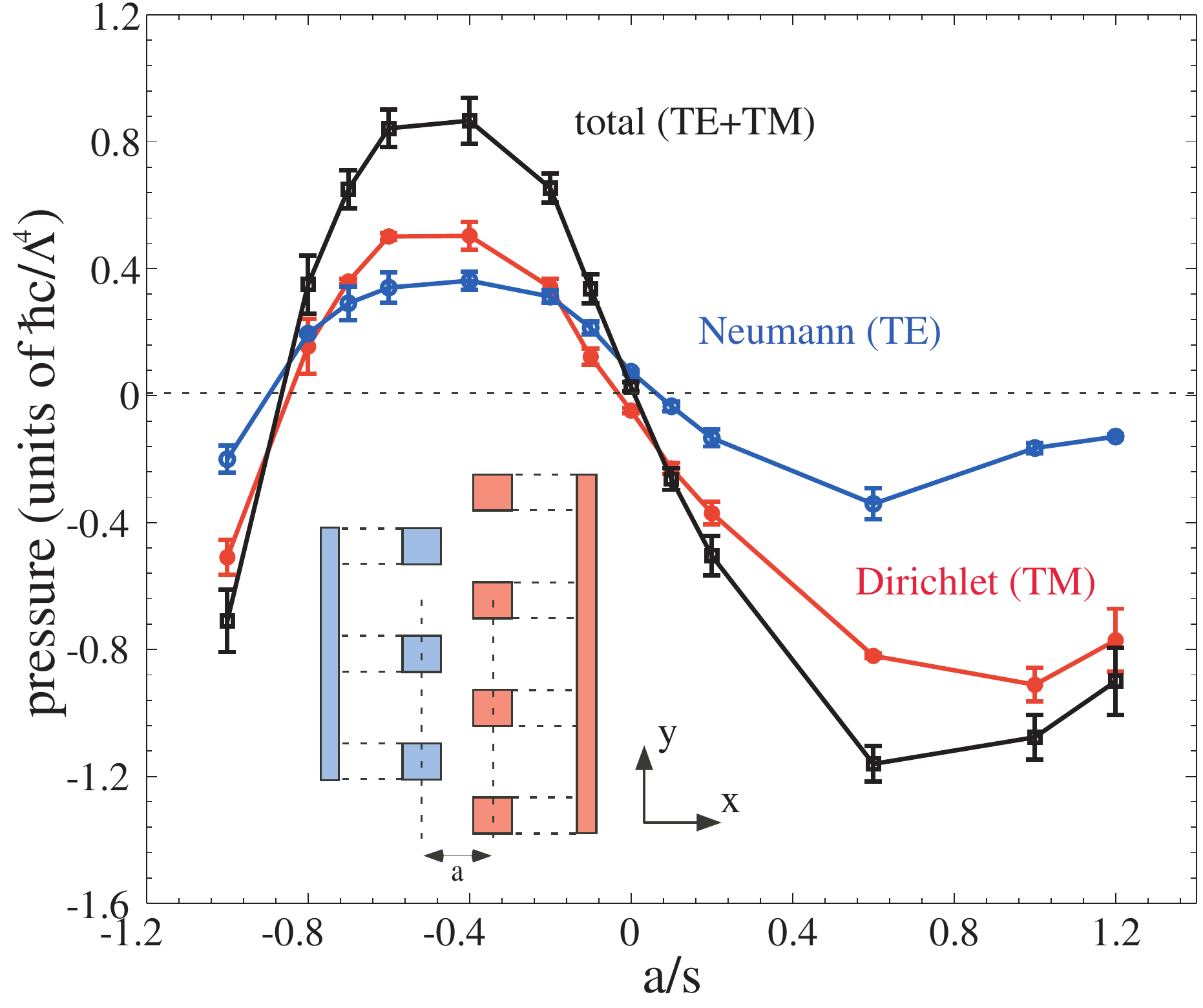}
\includegraphics[width=1.0\columnwidth]{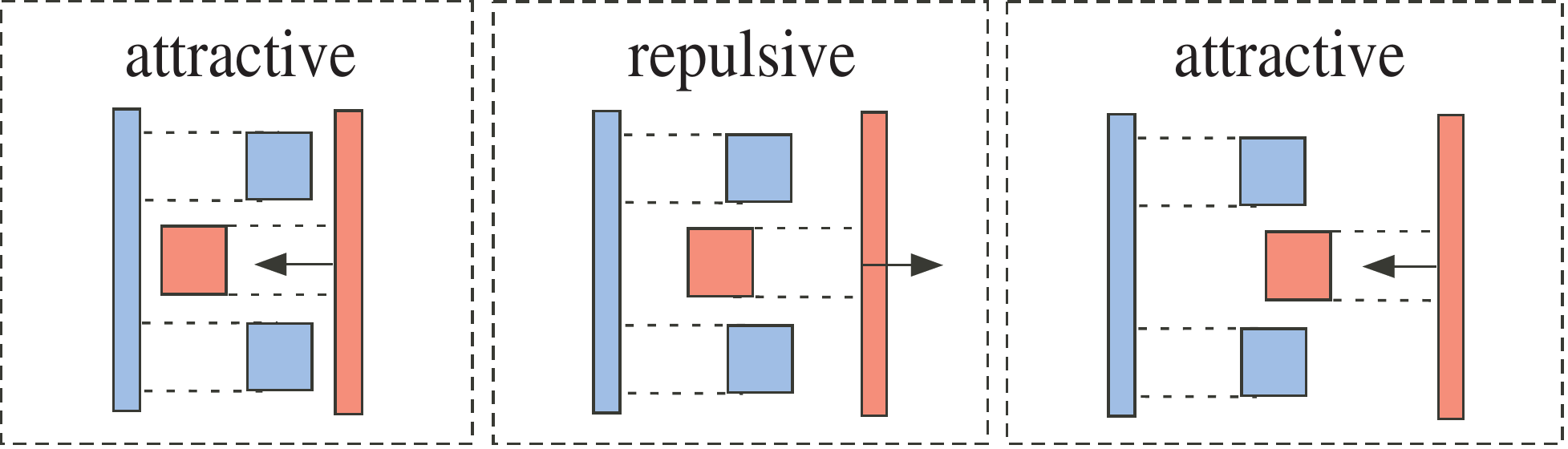}
\caption{(Color) Top: Plot of the Neumann (blue, TE), Dirichlet (red,
TM) and total (black, TE+TM) Casimir pressure (in units of $\hbar c /
\Lambda^4$) between the objects of \figref{geom}, as a function of
$a/s$. The inset illustrates a two-dimensional cross-section. Bottom:
Schematic indicating the various qualitatively different Casimir force
regimes between the two structures.}
\label{fig:plot}
\end{figure}

The geometry that we consider is depicted schematically in
\figref{geom}: we have two periodic sequences of metal ``brackets''
attached to parallel metal plates, which are brought into close
proximity in an interlocking ``zipper'' fashion.  In \figref{geom}, we
have colored the two plates/brackets red and blue to distinguish them,
but they are made of the same metal material.  This structure is not
mirror symmetric (and in fact is glide-symmetric, although the glide
symmetry is not crucial), so it is not required to have an attractive
Casimir force by \citeasnoun{KennethKl06}.  Furthermore, the structure
is connected and the objects can be separated via a rigid motion
parallel to the force (a consideration that excludes interlocking
``hooks'' and other geometries that trivially give repulsive
forces). This structure is best understood by considering its
two-dimensional cross-section, shown in \figref{geom}(right) for the
middle of the brackets: in this cross-section, each bracket appears as
an $s\times s$ square whose connection to the adjacent plate occurs
out-of-plane.  (Here, the brackets are repeated in each plate with
period $\Lambda = 2s+2h$ and are separated from the plates by a
distance $d$.  The plates are separated by a distance $2d+s+a$, so
that $a=0$ is the point where the brackets are exactly aligned.)  The
motivation for this geometry is an intuitive picture of the Casimir
force as an attractive interaction between surfaces.  When the plates
are far apart and the brackets are not interlocking, the force should
be the ordinary attractive one.  As the plates move closer together,
the force is initially dominated by the attractions between adjacent
bracket squares, and as these squares move past one another ($a < 0$
in \figref{geom}), one might hope that their attraction leads to a net
repulsive force pushing the plates apart.  Finally, as the plates move
even closer together, the force should be dominated by the
interactions between the brackets and the opposite plate, causing the
force to switch back to an attractive one.  This intuition must be
confirmed by an exact numerical calculation, however, because actual
Casimir forces are not two-body attractions and can sometimes exhibit
qualitatively different behaviors than a two-body model might predict~\cite{Zaheer07}.  Such a computation of the total force per
unit area is shown in \figref{plot}, and demonstrates precisely the
expected sign changes in the force for the three separation regimes.
These results are discussed in greater detail below.

Previous theoretical studies of Casimir forces in geometries with
strong curvature have considered a variety of objects and shapes.
Forces between isolated spheres~\cite{Emig07} and isolated
cylinders~\cite{RahiRo07}, or between a single sphere~\cite{Bordag06},
or cylinder~\cite{Bordag06, emig06} and a metal plate, all exhibit
attractive forces that decrease monotonically with separation.  When a
pair of squares~\cite{Rodriguez07:PRL} or cylinders~\cite{RahiRo07}
interacts in the presence of two adjacent metal sidewalls, the force
is still attractive and monotonic in the square/square or
cylinder/cylinder separation, but is a nonmonotonic function of the
sidewall separation.  When two corrugated surfaces are brought
together in a way that breaks mirror symmetry (i.e., the corrugations
are not aligned between the two surfaces), a lateral force can
arise~\cite{emig03_1, Rodrigues06:torque}, and an asymmetric lateral
force from asymmetric corrugations can lead to a ``ratchet'' effect in
which random forces preferentially displace the plates in one
direction~\cite{Emig07:ratchet}. Such a lateral force has also been
observed experimentally~\cite{Mohideen02:lateral}.  In the geometry of
\figref{geom}, in contrast, there is no lateral force (due to a
mirror-symmetry plane perpendicular to the plates), and hence we
consider only the normal force between the plates.  Because of the
strong curvature of the surfaces relative to their separations, simple
parallel-plate approximations are not valid (although we consider
their qualitative accuracy below), and the force must be computed
numerically.

The numerical method we employ is based on integration of the mean
stress tensor, evaluated in terms of the imaginary-frequency Green's
function via the fluctuation-dissipation
theorem~\cite{Rodriguez07:PRA}.  The Green's function can be evaluated
by a variety of techniques, but here we use a simple finite-difference
frequency-domain method~\cite{Rodriguez07:PRA,Christ87} that has the
advantage of being very general and simple to implement at the expense
of computational efficiency.  In particular, the computation involves
repeated evaluation of the electromagnetic Green's function,
integrated over imaginary frequency $w=-i\omega$ and a surface around
the object of interest.  The Green's function is simply the inverse of
a linear operator [$\nabla\times\nabla\times + w^2
\varepsilon(iw,\vec{r})$], which here is discretized using a
finite-difference Yee grid~\cite{Christ87} and inverted using the
conjugate-gradient method~\cite{bai00}. In order to simplify the
calculations, we assume the length of the brackets in the $z$
direction $L$ to be sufficiently long to make their contributions to
the force negligible (we estimate the minimum length below).  We can
therefore describe the geometry as both $z$-invariant and $y$-periodic
(with period $\Lambda$). This implies that it is only necessary to
compute the Green's function using an $xy$ unit cell, with the
periodic/invariant directions handled by integrating over the
corresponding wavevectors~\cite{Rodriguez07:PRA}.  Furthermore, we
approximate the bracket/plate materials by perfect metals, valid in the
limit of small lengthscales (which are dominated by long-wavelength
contributions where the skin depth is negligible).  In this case, the
contributions to the force can be separated into two polarizations:
transverse electric (TE) with the electric field in the $xy$ plane (a
scalar magnetic field with Neumann boundary conditions); and
transverse magnetic (TM) with the magnetic field in the $xy$ plane (a
scalar electric field with Dirichlet boundary
conditions)~\cite{Rodriguez07:PRA}, and these two contributions are
shown separately in \figref{plot}.

The resulting force per unit area between the plates, for the chosen
parameters $d/s=2$ and $h/s=0.6$, is plotted as a function of $a/s$ in
\figref{plot}~(Top); error bars show estimates of the numerical
accuracy due to the finite spatial resolution.  A number of unusual
features are readily apparent in this plot. First, the sign of the
force changes not only once, but twice. The corresponding zeros of the
force lie at $a/s \approx -0.8$ and $a/s \approx -10^{-2}$. The first
zero, $a/s \approx -0.8$, is a point of unstable equilibrium, to the
left of which the force is attractive and to the right of which the
force is repulsive. The second zero at $a/s \approx -10^{-2}$
corresponds to a point of stable equilibrium, with respect to
perpendicular displacements, for which the force is attractive to the
right and repulsive to the left.  (This point is still unstable with
respect to lateral displacements, parallel to the plates and
perpendicular to the brackets, however: any such lateral displacement
will lead to a lateral force that pulls the red and blue brackets
together.) In between these equilibria, the repulsive force has a
local maximum at $a/s \approx -0.5$.  Finally, at $a/s \approx 0.6$
the magnitude of the attractive force reaches a local maximum (a local
minimum in the negative force on the plot), and then decreases
asymptotically to zero as $a/s \rightarrow \infty$. Thus, as the two
objects move apart from one another, the force between them varies in
a strongly nonmonotonic fashion (distinct from the nonmonotonic
dependence on an external parameter shown in our previous
work~\cite{Rodriguez07:PRL, Zaheer07, RahiRo07}).  These three
different sign regimes are shown schematically in
\figref{plot}~(Bottom), as predicted by the intuitive picture
described above.

Since the qualitative features of the Casimir force in this geometry
correspond to the prediction of an intuitive model of pairwise surface
attractions, it is reasonable to ask how such a model compares
quantitatively with the numerical results.  The most common such
model is the proximity-force approximation (PFA), which treats the
force as a summation of simple ``parallel-plate''
contributions~\cite{Bordag06}.  (Another pairwise power-law heuristic
is the ``Casimir-Polder interaction'' approximation, strictly valid
only in the limit of dilute media~\cite{Tajmar04}.)  Applied to a
geometry with strong curvature and/or sharp corners such as this one,
PFA is an uncontrolled approximation and its application is
necessarily somewhat {\it ad hoc} (due to an arbitrary choice of which
points on the surfaces to treat as ``parallel plates''), but it
remains a popular way to quantify the crude intuition of Casimir
forces as pairwise attractions.

\begin{figure}[ht]
\includegraphics[width=1.0\columnwidth]{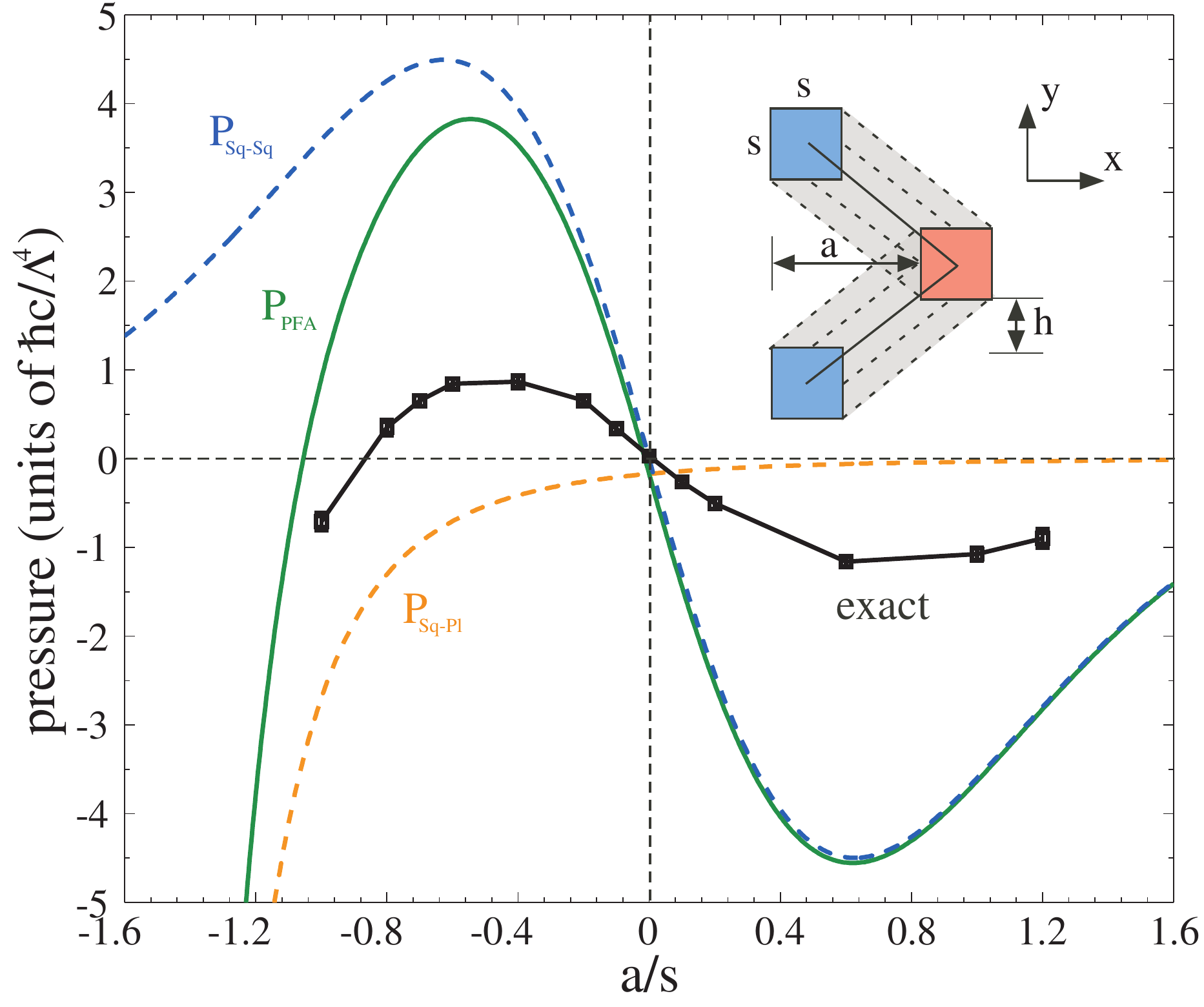}
\caption{(Color) Comparison of Casimir pressure (in units of $\hbar c
/ \Lambda^4$) as a function of $a/s$ between the stress-tensor (exact)
numerical results (black squares) and the proximity-force
approximation (solid green). Also shown are the individual
square--square (dashed blue) and square--plate (dashed orange)
contributions to the PFA force. Inset: Schematic illustration of the
chosen PFA ``lines of interaction'' between squares (dashed black
lines).}
\label{fig:plot2}
\end{figure}

Applying the PFA approximation to the two objects in \figref{geom}, we
treat the net force as a sum of three contributions: the force between
the two parallel plates, the force between each square and the
opposite plate, and the force between adjacent red and blue squares.
Namely,
\begin{equation} P_{\mathrm{PFA}} = \frac{1}{\Lambda L} \left(F_{\mathrm{pl-pl}} +
2 F_{\mathrm{sq-pl}} + 2 F_{\mathrm{sq-sq}}\right),
\label{eq:pfa}
\end{equation}
where the first term is the pressure between two parallel plates
(pl--pl), and the two remaining terms correspond to the square--plate
(sq--pl) and square--square (sq--sq) interactions. The factors of
$\Lambda$ and $L$ are introduced because these expressions are
computed per unit length in the $z$ direction, and per period in the
$y$ direction.

The first two PFA contributions are relatively simple to calculate
because they are between parallel metal surfaces, and thus (in the PFA
approximation) are the ordinary Casimir force weighted by the
respective areas:
\begin{eqnarray}
  P_{\mathrm{pl-pl}} &=& -\frac{\hbar c \pi^2 h}{120
  \Lambda}\frac{1}{(2d+a+s)^4} \\ P_{\mathrm{sq-pl}} &=& -\frac{\hbar
  c \pi^2 s}{240\Lambda} \frac{1}{(d+a)^4}
\end{eqnarray}
Computing the square--square force is less straightforward, since
there is some ambiguity as to what the PFA approximation even means
for two non-parallel surfaces (separate from the question of its
accuracy).  In PFA, one adds up ``parallel plate'' contributions to
the force between two objects by including a force between each point
on one surface and a corresponding point on the other surface, where
corresponding points are connected by parallel ``lines of
interaction.''  In this geometry, we take the lines of interaction to
lie parallel to the center-to-center displacement between two squares,
as depicted by the inset in \figref{plot2}, but of course this choice
is somewhat arbitrary.  (A similar choice was made by
\citeasnoun{Onofrio04} to define the PFA force between two eccentric
cylinders.)  The PFA force between one pair of squares is then:
\begin{multline}
    P_{\mathrm{sq-sq}} = 
\\ -\frac{\hbar c \pi^2 a}{240 \Lambda D^5}
    \left\{ \left[\frac{2|a|}{3}\left(H^3-1\right) + \frac{s H^3}{h}
    (Hh-|a|)\right]\Theta(Hh-a) \right. 
\\ +
    \left. \left[\frac{2Hh}{3}\left(A^3-1\right) + \frac{s
    A^3}{|a|-s}\left(|a|-Hh\right)\right] \Theta(|a|-Hh)\right\}
\end{multline}
where $D \equiv \sqrt{a^2+(h+s)^2}$, $H \equiv 1+s/h$ and $A \equiv
1-s/|a|$. The resulting net force is shown in \figref{plot2}, along
with the contributions due to the isolated square--square and
square--plate PFA forces (a separate line for the plate--plate
contributions is not shown because this contribution is always very
small).

For comparison, \figref{plot2} also shows the exact total force from
\figref{plot}, and it is clear that, while PFA captures the
qualitative behavior of the oscillating force sign, in quantitative
terms it greatly overestimates the magnitude of the repulsive force.
Of course, since it is an uncontrolled approximation in this regime
there is no reason to expect quantitative accuracy, but the magnitude
of the error illustrates how different the true Casimir force is from
this simple estimate.  The PFA estimate for the square--plate force,
however, does help us to understand one feature of the exact result.
If there were no plates, only squares, then the force would be zero by
symmetry exactly at $a=0$, and indeed the exact result including the
plates has zero force at $a \approx 0$; clearly, the contribution to
the force from the plates is negligible for $a \approx 0$, and this is
echoed by the PFA $P_{\mathrm{sq-pl}}$ force.  Also, using a PFA
approximation, one can attempt to estimate the order of magnitude of
the force contribution from the ends of the bracket, which was
neglected in the exact calculation.  This contribution to the total
force must decrease as $\sim 1/L$ for a fixed $a$, and is estimated to
be less than 1\% of the peak repulsive force for $L \gtrsim 60
\Lambda$.

Because the basic explanation for the sign changes in the force for
this structure is fundamentally geometrical, we expect that the
qualitative behavior will be robust in the face of imperfect metals,
surface roughness, and similar deviations from the ideal model here.
The main challenge for an experimental realization (for example, to
obtain a mechanical oscillator around the equilibrium point) would
appear to be maintaining a close parallel separation of the brackets
(although it may help that in at least one direction this parallelism
is a stable equilibrium).  Furthermore, although in this paper we
demonstrated one realization of a geometry-based repulsive Casimir
force, this opens the possibility that future work will reveal similar
phenomena in many other geometries.

We are grateful to M. Ibanescu for useful discussions.  This work was
supported in part by a U.~S. Department of Energy Computational
Science Graduate Fellowship under grant DE--FG02-97ER25308.


\begin{thebibliography}{10}

\bibitem{casimir}
H.~B.~G. Casimir, ``On the attraction between two perfectly conducting
  plates,'' {\em Proc. K. Ned. Akad. Wet.}, vol.~51, pp.~793--795, 1948.

\bibitem{moh3}
B.~W. Harris, F.~Chen, and M.~U., ``Precision measurement of the {Casimir}
  force using gold surfaces,'' {\em Phys. Rev.~A}, vol.~62, p.~052109, 2000.

\bibitem{decca2}
R.~S. Decca, F.~Fischbach, G.~L. Klimchitskaya, D.~E. Krause, D.~Lopez, and
  V.~M. Mostepanenko, ``Improved tests of extra-dimensional physics and thermal
  quantum field theory from new {Casimir} force measurements,'' {\em Phys.
  Rev.~D}, vol.~68, p.~116003, 2003.

\bibitem{KennethKl06}
O.~Kenneth and I.~Klich, ``Opposites attract: A theorem about the {Casimir}
  force,'' {\em Phys. Rev. Lett.}, vol.~97, p.~160401, 2006.

\bibitem{Kenneth02}
O.~Kenneth, I.~Klich, A.~Mann, and M.~Revzen, ``Repulsive {Casimir} forces,''
  {\em Phys. Rev. Lett.}, vol.~89, no.~3, p.~033001, 2002.

\bibitem{ShaoZh06}
C.-G. Shao, D.-L. Zheng, and J.~Luo, ``Repulsive {Casimir} effect between
  anisotropic dielectric and permeable plates,'' {\em Phys. Rev.~A}, vol.~74,
  p.~012103, 2006.

\bibitem{Boyer74}
T.~H. Boyer, ``{Van} der {Waals} forces and zero-point energy for dielectric
  and permeable materials,'' {\em Phys. Rev.~A}, vol.~9, pp.~2078--2084, 1974.

\bibitem{Munday07}
J.~N. Munday and F.~Capasso, ``Precision measurement of the {Casimir-Lifshitz}
  force in a fluid,'' {\em Phys. Rev.~A}, vol.~75, p.~060102(R), 2007.

\bibitem{Leonhardt07}
U.~Leonhardt and T.~G. Philbin, ``Quantum levitation by left-handed
  metamaterials,'' {\em New Journal of Physics}, vol.~9, no.~254, pp.~1--11,
  2007.

\bibitem{Sherkunov05}
Y.~Sherkunov, ``Van der {Waals} interaction of excited media,'' {\em Phys.
  Rev.~A}, vol.~72, p.~052703, 2005.

\bibitem{Emig07:ratchet}
T.~Emig, ``Casimir-force-driven ratchets,'' {\em Phys. Rev. Lett.}, vol.~98,
  p.~160801, 2007.

\bibitem{Rodriguez07:PRL}
A.~Rodriguez, M.~Ibannescu, D.~Iannuzzi, F.~Capasso, J.~D. Joannopoulos, and
  S.~G. Johnson, ``Computation and visualization of {Casimir} forces in
  arbitrary geometries: {Non-monotonic} lateral-wall forces and failure of
  proximity force approximations,'' {\em Phys. Rev. Lett.}, vol.~99, no.~8,
  p.~080401, 2007.

\bibitem{Rodriguez07:PRA}
A.~Rodriguez, M.~Ibannescu, D.~Iannuzzi, J.~D. Joannopoulos, and S.~G. Johnson,
  ``Virtual photons in imaginary time: computing {Casimir} forces in arbitrary
  geometries via standard numerical electromagnetism,'' {\em Phys. Rev.~A},
  vol.~76, no.~3, p.~032106, 2007.

\bibitem{Gies03}
H.~Gies, K.~Langfeld, and L.~Moyaerts, ``{Casimir} effect on the worldline,''
  {\em J. High Energy Phys.}, vol.~6, p.~018, 2003.

\bibitem{emig06}
T.~Emig, R.~L. Jaffe, M.~Kardar, and A.~Scardicchio, ``{Casimir} interaction
  between a plate and a cylinder,'' {\em Phys. Rev. Lett.}, vol.~96, p.~080403,
  2006.

\bibitem{Emig07}
T.~Emig, N.~Graham, R.~L. Jaffe, and M.~Kardar, ``Casimir forces between
  arbitrary compact objects,'' {\em Phys. Rev. Lett.}, vol.~99, p.~170403,
  2007.

\bibitem{Zaheer07}
S.~Zaheer, A.~W. Rodriguez, S.~G. Johnson, and R.~L. Jaffe,
  ``Optical-approximation analysis of sidewall-spacing effects on the force
  between two squares with parallel sidewalls,'' {\em Phys. Rev.~A}, vol.~76,
  no.~6, p.~063816, 2007.

\bibitem{RahiRo07}
S.~J. Rahi, A.~W. Rodriguez, T.~Emig, R.~L. Jaffe, S.~G. Johnson, and
  M.~Kardar, ``Nonmonotonic effects of parallel sidewalls on casimir forces
  between cylinders,'' {\em Phys. Rev.~A}, 2008.
\newblock In press.

\bibitem{Bordag06}
M.~Bordag, ``{Casimir} effect for a sphere and a cylinder in front of a plane
  and corrections to the proximity force theorem,'' {\em Phys. Rev.~D},
  vol.~73, p.~125018, 2006.

\bibitem{emig03_1}
T.~Emig, A.~Hanke, R.~Golestanian, and M.~Kardar, ``Normal and lateral
  {Casimir} forces between deformed plates,'' {\em Phys. Rev.~A}, vol.~67,
  p.~022114, 2003.

\bibitem{Rodrigues06:torque}
R.~B. Rodrigues, P.~A. Maia~Neto, A.~Lambrecht, and S.~Reynaud,
  ``Vacuum-induced torque between corrugated metallic plates,'' {\em Europhys.
  Lett.}, vol.~75, no.~5, pp.~822--828, 2006.

\bibitem{Mohideen02:lateral}
F.~Chen, U.~Mohideen, G.~L. Klimchitskaya, and V.~M. Mostepanenko,
  ``Demonstration of the lateral {Casimir} force,'' {\em Phys. Rev. Lett.},
  vol.~88, p.~101801, 2002.

\bibitem{Christ87}
A.~Christ and H.~L. Hartnagel, ``Three-dimensional finite-difference method for
  the analysis of microwave-device embedding,'' {\em IEEE Trans. Microwave
  Theory Tech.}, vol.~35, no.~8, pp.~688--696, 1987.

\bibitem{bai00}
Z.~Bai, J.~Demmel, J.~Dongarra, A.~Ruhe, and H.~Van Der~Vorst, {\em Templates
  for the Solution of Algebraic Eigenvalue Problems: A Practical Guide}.
\newblock Philadelphia: SIAM, 2000.

\bibitem{Tajmar04}
M.~Tajmar, ``Finite element simulation of {Casimir} forces in arbitrary
  geometries,'' {\em Intl. J. Mod. Phys. C}, vol.~15, no.~10, pp.~1387--1395,
  2004.

\bibitem{Onofrio04}
D.~A.~R. Dalvit, F.~C. Lombardo, F.~D. Mazzitelli, and R.~Onofrio, ``Casimir
  force between eccentric cylinders,'' {\em Europhys. Lett.}, vol.~67, no.~4,
  pp.~517--523, 2004.

\end{thebibliography}

\end{document}